\documentclass[conference]{IEEEtran}
\IEEEoverridecommandlockouts
\usepackage{cite}
\usepackage{amsmath,amssymb,amsfonts}
\usepackage{algorithmic}
\usepackage{graphicx}
\usepackage{textcomp}
\usepackage{xcolor}
\usepackage{float}
\usepackage{ifthen}
\usepackage{hyperref}

\definecolor{softgreen}{RGB}{44, 130, 76} % rich green, slightly muted
\def\BibTeX{{\rm B\kern-.05em{\sc i\kern-.025em b}\kern-.08em
    T\kern-.1667em\lower.7ex\hbox{E}\kern-.125emX}}

\newcommand{\authstyle}{2} % Set to 1 or 2

\begin{document}

\title{Identifying Protein Co-regulatory Network Logic by Solving B-SAT Problems through Gate-based Quantum Computing}

% === Authorship style 1 ===
\ifthenelse{\authstyle=2}{

\author{\IEEEauthorblockN{Aspen Erlandsson Brisebois}
\IEEEauthorblockA{\textit{Vaccine and Infectious Disease Organization (VIDO)} \\
\textit{University of Saskatchewan}\\
Saskatoon, Canada \\
\textit{Engineering Science, Faculty of Applied Science}\\
\textit{University of Toronto}\\
Toronto, Canada\\
gmr482@mail.usask.ca}\\~~\\
\IEEEauthorblockN{Zahed Khatooni}
\IEEEauthorblockA{\textit{Vaccine and Infectious Disease Organization (VIDO)} \\
\textit{University of Saskatchewan}\\
Saskatoon, Canada \\
zahed.khatooni@usask.ca}\\~~\\
\IEEEauthorblockN{Steven Rayan}
\IEEEauthorblockA{\textit{Centre for Quantum Topology and Its Applications (quanTA)} \\
\textit{Department of Mathematics and Statistics}\\ 
\textit{University of Saskatchewan}\\
Saskatoon, Canada \\
rayan@math.usask.ca}
\and
\IEEEauthorblockN{Jason Broderick}
\IEEEauthorblockA{\textit{Vaccine and Infectious Disease Organization (VIDO)} \\
\textit{University of Saskatchewan}\\
Saskatoon, Canada \\
\textit{Engineering Science, Faculty of Applied Science}\\
\textit{University of Toronto}\\
Toronto, Canada\\
xoq037@mail.usask.ca}\\~~\\
\IEEEauthorblockN{Heather L. Wilson}
\IEEEauthorblockA{\textit{Vaccine and Infectious Disease Organization (VIDO)} \\
\textit{University of Saskatchewan}\\
Saskatoon, Canada \\
heather.wilson@usask.ca}\\~~\\
\IEEEauthorblockN{Gordon Broderick}
\IEEEauthorblockA{\textit{Vaccine and Infectious Disease Organization (VIDO)} \\
\textit{Department of Mathematics and Statistics}\\
\textit{Centre for Quantum Topology and Its Applications (quanTA)} \\
\textit{University of Saskatchewan}\\
Saskatoon, Canada \\
awr794@mail.usask.ca}
}

}{

% === Authorship style 2 ===

\author{
    \IEEEauthorblockN{
        Aspen Erlandsson Brisebois\textsuperscript{1,2}, 
        Jason Broderick\textsuperscript{1,2}, 
        Zahed Khatooni\textsuperscript{1}, 
        Heather L. Wilson\textsuperscript{1},\\
        Steven Rayan\textsuperscript{3,4}, 
        Gordon Broderick\textsuperscript{1,3,4}
    }\\

    \IEEEauthorblockA{\textsuperscript{1} Vaccine and Infectious Disease Organization (VIDO), University of Saskatchewan, Saskatoon, Canada}
    \IEEEauthorblockA{\textsuperscript{2} Engineering Science, Faculty of Applied Science, University of Toronto, Toronto, Canada}
    \IEEEauthorblockA{\textsuperscript{3} Department of Mathematics and Statistics, University of Saskatchewan, Saskatoon, Canada}
    \IEEEauthorblockA{\textsuperscript{4} Centre for Quantum Topology and Its Applications (quanTA), University of Saskatchewan, Saskatoon, Canada}
}

}

\maketitle

\begin{abstract}
There is growing awareness that the success of pharmacologic interventions on living organisms is significantly impacted by context and timing of exposure. In turn, this complexity is leading to an increasing focus on regulatory network dynamics in biology and our ability to represent these at a high level of fidelity, \textit{in silico}. Logic network models continue to show great promise in this domain and their parameter estimation can be formulated as a constraint satisfaction problem (CSP) that is especially well-suited for the often sparse and incomplete data associated with biology. Unfortunately, even in the case of Boolean logic, the combinatorial complexity of these problems grows rapidly, challenging our ability to create such models at physiologically-relevant scales. That said, quantum computing, while still nascent, facilitates novel information-processing paradigms with the potential for transformative impact in problems such as this one. In this work, we take a first step at actualizing this potential by identifying the structure and Boolean decisional logic of a well-studied regulatory network linking five proteins involved in the neural development of the mammalian cortical area of the brain. We identify the protein-protein connectivity and binary decisional logic governing this network by formulating it as a Boolean Satisfiability (B-SAT) problem. We then employ Grover's algorithm to solve the NP-hard problem faster than the exponential time complexity required by deterministic classical algorithms. Using approaches deployed on both quantum simulators and actual noisy intermediate scale quantum (NISQ) hardware, we accurately recover several high-likelihood models from very sparse protein expression data. The results highlight the differential roles of various data types in supporting accurate models; the impact of quantum algorithm design as it pertains to the mutability of quantum hardware, and the opportunities for accelerated discovery enabled by this approach.
\end{abstract}


\textbf{\emph{Note on Authorship}}---\textbf{AEB and JB are joint first authors; GB, SR, and HLW are joint senior authors.}

\section{Introduction}
Agility in response, broad adaptability, and robustness in function in the face of environmental challenges and internal noise are fundamental characteristics of biological systems. The importance of individual biological components notwithstanding, this robust adaptability arises in large part not from a simple accumulation of independent actions but from complex and coordinated interactivity between these components \cite{blanchini_2011} \cite{silva_rocha_2010}. Logic circuit models \cite{abou_jaoude_2016} have provided a highly useful mathematical formalism with which to better understand the roles of network structure, regulatory programming and control dynamics in ensuring the functional properties of living systems \cite{alon_2007}. In an extension of foundational work by Thomas 1998 \cite{thomas_1998}, our group has focused on the assembly of regulatory network models from prior knowledge, using experimental data to constrain the choice of model rather than construct the model \textit{de novo}. This shift from discovery-driven to hypothesis-driven model assembly has made it possible to effectively leverage sparse, partially observed and incomplete data that would otherwise pose a significant challenge to conventional data-mining techniques. In essence, admissible network variants and sets of decisional logic settings are selected such that the predicted network dynamics do not contradict available experimental observations and expected behaviors (i.e. dynamic stability). These sets of parameter values are found by formulating and solving a large constraint satisfaction (SAT) problem \cite{sedghamiz_2019}. The size of the resulting combinatorial optimization problem grows exponentially with the number of network entities and relationships significantly limiting the size of the biological system and fidelity with which it can be represented. From cryptanalysis to protein design, efficient solutions to SAT problems of this type are of broad interest. Advances in algorithmic approaches like conflict-driven clause learning (CDCL) \cite{silva_1996} attempt to reduce the average complexity using certain heuristics to parse the expression, and more recently harness machine learning techniques \cite{guo_2023}, to produce more efficient search strategies.  Despite these advances, broadly applicable classical polynomial-time algorithms appear limited to runtime complexity of \(\mathcal{O}(2n)\). Quantum computing, by virtue of its fundamental physical properties, namely superposition of states and entanglement, constitutes a very promising alternative technology for more efficiently solving complex problems such as this with very large search spaces. 

In this work, our primary focus is to explore the practical feasibility of applying quantum computing in the field to reverse engineer immune response networks at a scale that is immediately useful in advancing the network-informed design of immune modulatory interventions, both preventative and restorative. Towards this, we use as our model a well-studied biological regulatory network of morphogens/signaling proteins that are expressed during brain development and that help establish the expression patterns of transcription factors (TFs). Specifically, Fgf8 is a signaling protein essential in establishing the midbrain-hindbrain boundary \cite{harada_2016}\cite{lee_1997}. It controls the graded expression of TFs such as COUP-TFI, Emx2, Pax6, and Sp8 that help regulate expressions of genes whose gene products (i.e. proteins) convey positional or area identities to cortical progenitors \cite{oleary_2007}. We use the corresponding regulatory logic model proposed by \cite{giacomantonio_2010} to generate simulated time course instances then attempt to use quantum algorithms to reverse engineer the original regulatory logic from these simulated data points . In keeping with our previous work, this problem was articulated as a Boolean Satisfiability problem (B-SAT) to facilitate the design of a corresponding quantum circuit. Specifically, the regulatory logic governing state transition of each network node was stated as a Boolean expression where the solution consists of a set of logical operators and upstream regulators that render the expression True. At its core, SAT remains a search problem, where the choice of an expression \(f(s_1, s_2, ... , s_n)\) must match specific instances.  Grover's algorithm is framed in this exact way, with an arbitrary function enclosed in an 'oracle' and a search space of \(2n\) possible values. Using Grover’s algorithm and the principles of quantum superposition of states, the complexity of the search was reduced to \(\mathcal{O}(\sqrt{n})\) and the expression corresponds to a logic statement governing state transition.  To solve this logic estimation problem, we leveraged the implementation of Grover’s algorithm and the oracle synthesizer available under the Qiskit Python programming environment \cite{javadi_abhari_2024} to generate the corresponding quantum circuit.  Further reductions in the complexity of the quantum circuit produced by the built-in Qiskit oracle synthesizer were obtained by carefully introducing additional problem-specific information. We show that both the manually designed circuit assembled using low-level functions and the optimized circuit leveraging high-level functions correctly recover the true logic in a noiseless quantum simulation environment when determinacy is supported. As biomedical data is invariably time-consuming, resource intensive, invasive and often risky to collect, the reverse engineering of these biological networks from very sparse data is of primary importance. Interestingly, when presented with an underdetermined problem where multiple competing solutions legitimately exist, we find that actual NISQ hardware recovers the true logic with a higher ranking than that obtained under noiseless conditions.  Likewise, we also find that the use of low-level functions resulting in a lower instruction count but increased qubit usage and potentially higher noise may deliver a more biologically relevant ranking highlighting the potentially useful role for noise inherent in these platforms to provide improved sampling of plausible solutions.

\section{Methods}

\subsection{A protein regulatory network}

The model used as a proof for this concept defines the state transition logic of five proteins for an arbitrary timestep: Fgf8, Emx2, Pax6, Sp8, and COUP-TFI. The equations are defined as follows:

{ \small
\begin{equation} \label{fgf8stateeq}
    \mbox{Fgf8}(t+1) = \mbox{Fgf8}(t) \land \lnot \mbox{Emx2}(t) \land \mbox{Sp8}(t)
\end{equation}
\vspace{-0.1in}
\begin{equation} \label{emx2stateeq}
    \mbox{Emx2}(t+1) = \lnot \mbox{Fgf8}(t) \land \lnot \mbox{Pax6}(t) \land \mbox{COUP-TFI}(t) \land \lnot \mbox{Sp8}(t)
\end{equation}
\vspace{-0.1in}
\begin{equation} \label{pax6stateeq}
    \mbox{Pax6}(t+1) = \lnot \mbox{Emx2}(t) \land \lnot \mbox{COUP-TFI}(t) \land \mbox{Sp8}(t)
\end{equation}
\vspace{-0.1in}
\begin{equation} \label{sp8stateeq}
    \mbox{Sp8}(t+1) = \mbox{Fgf8}(t) \land \lnot \mbox{Emx2}(t)
\end{equation}
\vspace{-0.1in}
\begin{equation} \label{couptfistateeq}
    \mbox{COUP-TFI}(t+1) = \lnot \mbox{Fgf8}(t) \land \lnot \mbox{Sp8}(t)
\end{equation}
}

\subsection{Assumptions and General Formulation of State Equations} \label{sec:assump}

We begin by formulating the SAT problem such that the parameter vector $x$ describes a particular logic model, which is itself a set of Boolean equations. A specific instance of the parameter vector $x$ is a solution if it defines a corresponding logic model that satisfies all the constraints imposed by the observed sample states manifested by the system. The first step was to choose a general structure for the logic model. We chose the general conjunctive normal form (CNF), since any Boolean state transition expression can be represented as the conjunction ($\land$) of one or more disjunctions ($\lor$) of one or more literals ($s_i$ or  $\lnot s_i$) corresponding to the current state of each network node ($S(t)$) (see Eqns. \ref{fgf8stateeq}-\ref{couptfistateeq}). However, even with only a handful of network state variables, the possibility of the same state variable interacting with more than one clause rapidly increases the number of possible solution statements. Thus, another simplification can be made which assumes that all clauses contain only one literal, the \textit{minterm canonical form}. This form is called a \textit{minterm} since it consists of the minimum number of possible literal assignments. There are $2^K$ minterms of $K$ (where $K=5$ proteins in this example) state variables $s$, since a variable in the minterm expression can only exist in its direct or its complemented form. By assuming a more specific form for the state equations, the problem is more easily parametrized and becomes a SAT problem with a large but reasonably finite search space, since variables can only appear once or not at all without introducing redundancy. Furthermore, we make the simplifying assumption that the systems to be solved are fully discretized in time and that the next state $S(t+1)$ of the system is a function of only its previous state $S(t)$. For example, we define the current state of a system $S(t)$ with $K$ state variables (nodes) as:

\begin{equation} \label{statevectort}
    S(t) = [s_1(t), s_2(t), s_3(t), \cdots]
\end{equation}

with the next state expressed as

\begin{equation} \label{statevectort1}
    S(t+1) = [s_1(t+1), s_2(t+1), s_3(t+1), \cdots]
\end{equation}

where

\begin{equation} \label{statefunction}
    S(t+1)=f(S(t))
\end{equation}

and where $i$ is the index of a state variable $i=1:K$. In formulating the SAT problem, there are two parameters for each system variable (i.e. $2K$), namely (i) inclusion of the literal $s_i(t)$ or (ii) the inclusion of its complement negation $\lnot s_i(t)$ using the NOT operator ($\lnot$).  By ranging over the full set of possible values for these two parameters for each variable, every possible minterm expression of state $S$ is representable. This state transition function $f(S(t))$ can be formally written in extended Baucus-Naur form (EBNF) and appears as follows:

\begin{equation} \label{baucusnaurform}
    f(S(t))=[\lnot s_1 (t)]|[s_1 (t)]\land[\lnot s_2 (t)]|[s_2 (t)]\land[\lnot s_i(t)]|[s_i(t)] \cdots
\end{equation}

Parametrizing the problem in this way effectively limits the complexity of modelling a $K$-variable network to an upper bound of $2K$ as opposed to the potentially much larger number of parameters necessary to fully define the general conjunctive normal form in which literals can appear more than once and in multiple different clauses. Though each clause has been reduced to a single literal, their combined use is sufficient to generate all of the state transition equations presented in \cite{weidner_2024}. In the case of more complex state transition dynamics where certain literals are suspected to be clauses of several literals instead of one, the latter can be formulated as component BSAT problems and solved independently. Importantly, deployment of this problem on current Noisy Intermediate Scale Quantum (NISQ) platforms implicitly requires that certain simplifying assumptions be made to accommodate the available quantum volume of the device, given in terms of the maximum quantum circuit width (number of qubits) and depth (number of gates).  A more complex model form describing network state transition could  be applied to this problem; that said, this compact representation not only adequately captures the dynamics of the system but also facilitates deployment on current NISQ hardware given current limitations.

\subsection{Mapping to Grover's Algorithm}

After creating a state representation defined by minterm expressions, we are left with the task of referring the problem to Grover’s algorithm, which accepts a single oracle function (i.e. Boolean expression) as input and solves for the bitstrings in a $K$-bit space that satisfy it (i.e. $f (S) = 1$). Since the next state of each system state variable $S$ is assumed to be defined by its own equation, the same approach described in this section can be applied to each variable individually, yielding a different oracle function. Using the description of the network system and state transition equations posed in Section \ref{sec:assump}, the search problem can be formulated as follows with Grover’s algorithm in mind:\\

\begin{enumerate}
    \item List the parameters that describe a given $K$-variable state transition equation

    \begin{equation} \label{fullstatefunc}
        S(t+1) = f (S(t)) = f (s_1(t), s_2(t), s_3(t), ...)
    \end{equation}

    For the equivalent minterm form, let us define the set of $2K$ paremeters

    \begin{equation} \label{params}
        \{s_1(t)_{\text{include}}, s_1(t)_{\text{not}}, s_2(t)_{\text{include}}, s_2(t)_{\text{not}}, \cdots \}
    \end{equation}

    \item Define the $2K$-bit search space in which the value of each logic parameter is determined by the value of a binary digit in a $2K$-bit string. For example, for two system variables—and four parameters—the set of bit strings would be

    \begin{equation} \label{bitstrings}
        \{0000, 0001, 0010, 0011, …, 1110, 1111\}	
    \end{equation}

    The size of this set grows exponentially with $K$ (it is precisely $2^{2K}$ elements in size).

    \item Collect samples that will constrain the behaviour of the state transition equation defined by the logic parameters. A sample is a pair that contains (i) the value of the state variable,  $f (S(t))$; and (ii) the set of values for each system variable that result in that state, $(s_1(t)^*, s_2(t)^*, s_3(t)^*, \cdots)$.

    \item Create an oracle function of the logic parameters 

    \begin{equation} \label{oraclefunc}
        O (s_1(t)_\text{include}, s_1(t)_\text{not}, s_2(t)_\text{include} , s_2(t)_\text{not} , ...)
    \end{equation}

    using the sampled network state $S^*(t)$; if $f (s_1(t)^*, s_2(t)^*, s_3(t)^*, ...) = S(t+1)$, then $\{s_1(t)_\text{include}, s_1(t)_\text{not}, s_2(t)_\text{include} , s_2(t)_\text{not} , \cdots \}$ should parametrize a minterm expression that includes $s_1(t)$, $s_2(t)$, $s_3(t)$, ... and equals $S(t+1)$ for

    \begin{equation} \label{paramassignments}
        s_1(t) = s_1(t)^*, s_2(t) = s_2(t)^*, s_3(t) = s_3(t)^*, ...
    \end{equation}

    For example, suppose that a two-variable system has a subsequent state of (1, 1) when the current state is (1, 0) and also assume that we are currently investigating the logic for $s_1(t)$. When $s_1(t) = 1$ and $s_2(t) = 0$, $s_1(t+1) = 1$, so the state transition equation would appear as:

    % \begin{equation} \label{stateeqexample1}
        $$s_1(t) \land \lnot s_2(t)$$
    % \end{equation}

    or

    % \begin{equation} \label{stateeqexample2}
        $$s_1(t)$$
    % \end{equation}

    Thus, $s_1(t)_\text{include}$ must be 1 and $s_1(t)_\text{not}$ must be 0. If $s_2(t)_\text{include}$ is 0, then the choice of $s_2(t)_{not}$ is arbitrary—since it is not included anyway. If $s_2(t)_\text{include}$ is 1, then $s_2(t)_\text{not}$ must also be 1. For this very simple example, $O$ must include the clauses

    \begin{equation} \label{exampmleclause1}
        (s_1(t)_\text{include} \land s_1(t)_\text{not} \land \lnot s_2(t)_\text{include})
    \end{equation}

    and

    \begin{equation} \label{exampleclause2}
        (s_1(t)_\text{include} \land s_1(t)_\text{not} \land s_2(t)_\text{include} \land s_2(t)_\text{not})
    \end{equation}

    \item Once the oracle function is defined using all available samples, solving for this satisfiability criteria produces a compliant regulatory logic.\\
    
\end{enumerate}

In this way, we converge towards equations that correspond with the allowed form of each variable’s state transition logic, and these are encoded sequentially one state variable at a time. The time complexity of solving this equation is exponential with $K$ for a classical computer, but using Grover’s algorithm to compute the solutions is logarithmic in $K$. By layering multiple known $S(t)$ input / $S(t+1)$ output sample pairs into the circuit for a common set of qubits, we incrementally create a more precise and deterministic oracle. More specifically, the oracle has been satisfiably defined if the corresponding parameter values define a logic model that explains all state transition constraint input/output pairs. The \textit{diffuser} is the next part of the quantum circuit, which translates the information encoded into the quantum phase of each qubit describing the logic parameter’s most likely value. Assessing the oracles identified over many iterations then serves to provide a distribution of the most probable logic models for the given system.

\subsection{Optimal  Number of Grover Algorithm Iterations}
\label{sec:optimal_iteration_number}

The optimal number of Grover iterations is essential for the efficient execution of Grover's algorithm. The algorithm relies on repeatedly applying the oracle-diffuser stage to iteratively rotate the quantum state vector toward the solution space. The expression for the optimal iteration count is given by:

\begin{equation} \label{optGroveriters}
    m = \left|\frac{4}{\pi \arcsin(\frac{t}{2^n})}\right|
\end{equation}

where $t$ is the number of solutions within the search space, and $n$ is the number of qubits used by the diffuser \cite{brassard_2002}. This formula critically depends on the knowledge of $t$, the number of solutions, which is often not readily available for most problems, including the one discussed here. Quantum amplitude estimation is a well-understood method that can determine $t$ efficiently. However, due to the current hardware limitations of NISQ devices amplitude estimation remains out of reach for the time being. Therefore, in this work we utilized classical methods to approximate $t$, allowing us to focus on the design and optimization of the oracle component. As quantum phase estimation implementations mature and quantum hardware improves, amplitude estimation will become a practical tool to compliment the Oracles developed in this paper, making the results entirely independent of classical computation.

\subsection{Implementing the Quantum Circuit with the Qiskit SDK}

We used two approaches to construct our quantum circuit, one consisting of a manual design relying on low-level functions, the other using high-level functions to assist and optimize the design. The first implementation consisted of a carefully constructed quantum circuit consisting of a custom Grover oracle that is generated procedurally gate-by-gate. Additional \textit{ancillary qubits}--qubits supplemental to the main $2K$ parameter qubits are used to the store intermediate quantum states after the application of each sample constraint throughout the course of the computation. To generate the Grover oracle, Python code utilizing low-level $Qiskit$ primitives (e.g. quantum registers, multiple-control / multiple-target gates) assembles the oracle gate-by-gate based on the values of $S(t)^*$ and $f(S(t)^*)$, the sample constraints. The Python program is designed to be case agnostic, and its generality is limited only by the assumptions discussed in Section \ref{sec:assump}. Solving for a different number of state variables or sampled state constraints is a matter of simply providing a different input sample data file. By design, the program will dynamically build a quantum circuit corresponding to the given input sample and the new SAT problem within the limitations of Section \ref{sec:assump} and the provided sample constraints.

In detail, this first approach involves the following steps:

\begin{enumerate}
    \item Define the variables that represent the Boolean model parameters discussed in Section \ref{sec:assump} and organize these into two separate \textit{quantum registers} (groups of qubits): one for the “include” parameters, and one for the “not” (complement) parameters.

    \item Designate ancillary qubits as a place to store the result of intermediate computations. Call the quantum register that encapsulates these qubits $T$. 

    \item Initialize the system by setting each qubit in $T$ to the value of the input nodes in the first sample constraint.

    \item Next, compute the result of \(s_{\text{include},i}\) and \(t_i\) for each \(i=1:K\) and store the result in \(t_i\) using a combination of \textit{CNOT} and \textit{Toffoli} gates.

    \item Then, apply a multi-control multi-target (MCMT) \textit{CNOT} gate, using the qubits in the \(T\) register as the control and target.

    \item Reinitialize all qubits to their initial state after each sample constraint and repeat steps (3)-(5) for each sample constraint provided. 

    \item Using a multi-control single-target (MCST) Z gate, which performs a conditional phase flip on the quantum state before applying the Grover diffuser and measuring the value of each \(s_\text{include}\) and \(s_\text{not}\) qubit.
\end{enumerate}

The measurements taken from the end indicate the values of the parameters--i.e. whether to include the variable in the minterm and whether to take its complement $\lnot$ that constructs a Boolean state equation that satisfies the constraints (as provided to the program that constructed the software description of the quantum circuit).

The second implementation leverages a built-in high-level library function in IBM’s Qiskit SDK which converts a high-level symbolic equation like the one in Section \ref{sec:assump} into a Grover oracle circuit directly. Leveraging the high-level functionality is quite straightforward: we need only define the general state transition logic equation for our specific model and given sample data, then use the \textit{PhaseOracle()} function from \textit{qiskit.circuit.library} and apply it to our Boolean expression containing the provided sample constraints. Optionally, a separate Boolean optimization library such as \textit{sympy} can be used to simplify the expression prior to passing it to the phase oracle generator. Implementation using this high-level function has the advantage of using fewer qubits since it omits the use of ancillaries, adhering more closely to the “all-in-place” style shown in most textbook representations of Grover’s algorithm. In brief, manual assembly is much more flexible and open to customization, leaving more circuit volume in which to optimize the configuration. However, this approach is also more time consuming and requires more attention to detail to be implemented correctly. In comparison, the function-assisted implementation currently executes with a faster runtime but offers virtually very little opportunity for additional optimization. 

\begin{figure}[h!]
    \centering
    \includegraphics[width=1\linewidth]{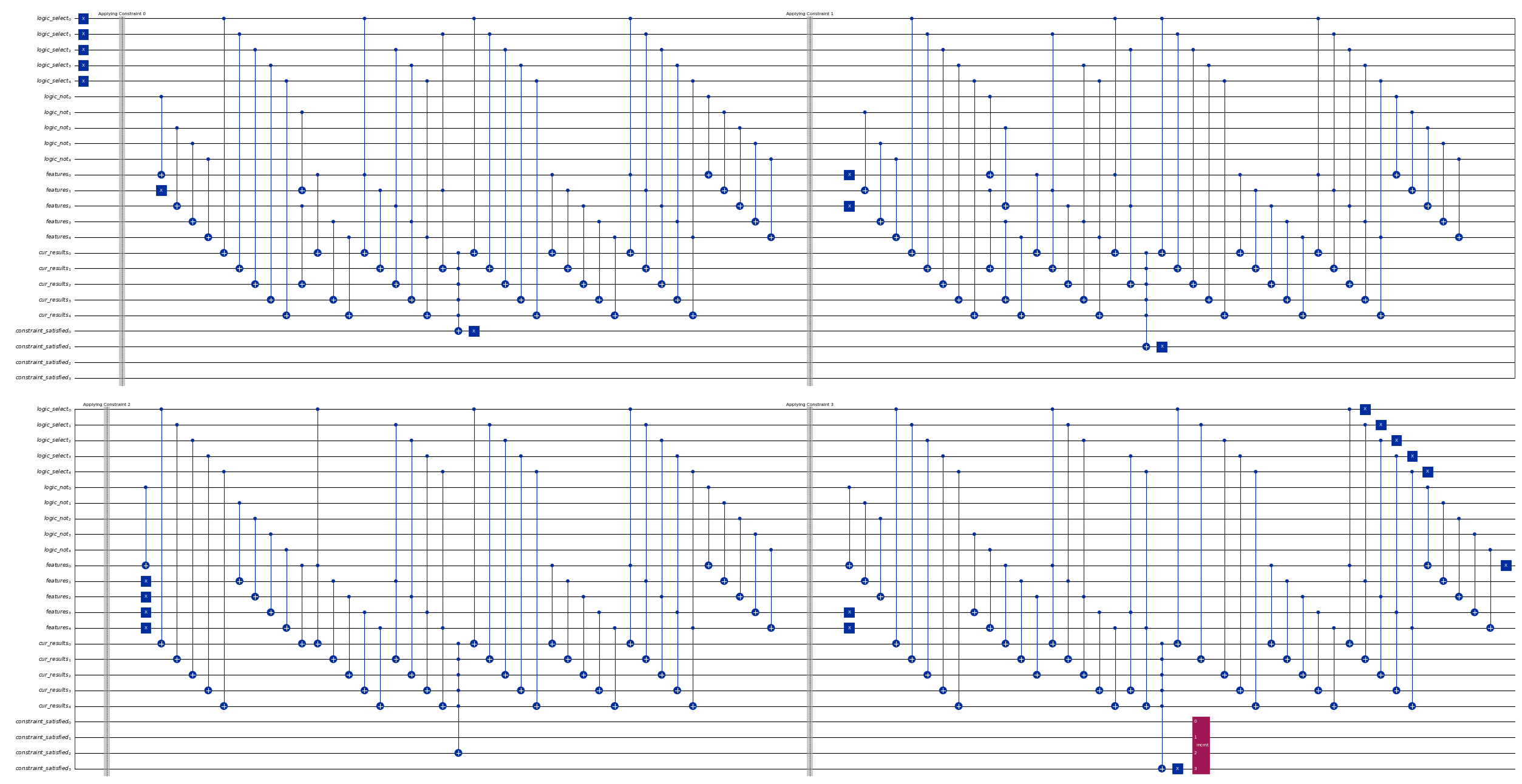}
    \caption{\textnormal{Handcrafted Oracle for four constraints applied (split across two lines) designed using low-level functions.}}
    \label{fig:handcrafted_circuit_rendering_4_constraints}
\end{figure}

\begin{figure}[h!]
    \centering
    \includegraphics[width=1\linewidth]{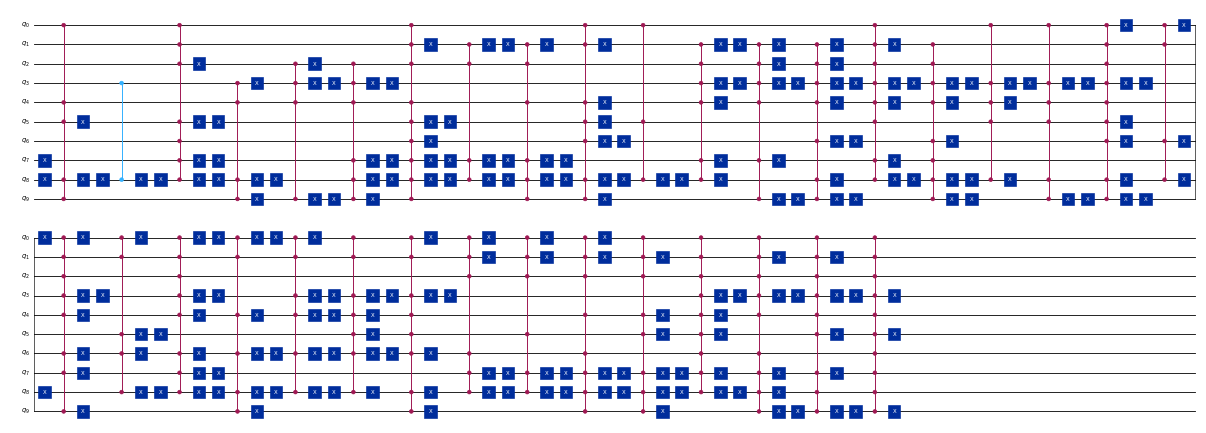}
    \caption{Phase oracle for four constraints designed using Qiskit high-level functions.}
    \label{fig:qiskit_generated_circuit_rendering_4_constraints}
\end{figure}

A rendering of the custom approach (1) and built-in oracle approach (2) are shown in Figure \ref{fig:handcrafted_circuit_rendering_4_constraints} and Figure \ref{fig:qiskit_generated_circuit_rendering_4_constraints}, respectively. Note that these figures show each circuit at different levels of abstraction; the custom-generated oracle may appear more complicated below because it was constructed gate-by-gate, as shown. Still, both oracles transpile into a similarly complex circuit as the circuit built using the \textit{Qiskit SDK}’s \textit{PhaseOracle()} function.

\section{Results}
\label{sec:results}

In the case of this example problem, both the manual circuit designed using low-level functions and the design assisted by high-level functions (Figure 1) produce the same results under noise-free quantum simulator conditions (Figure 2), consistent with expectations. Accordingly, we shall discuss the output produced using the high-level functions and compare runtime performance between both methods.

\subsection{Execution in a Noiseless Quantum Simulation Environment}

\begin{figure}[H]
    \centering
    \includegraphics[width=1\linewidth]{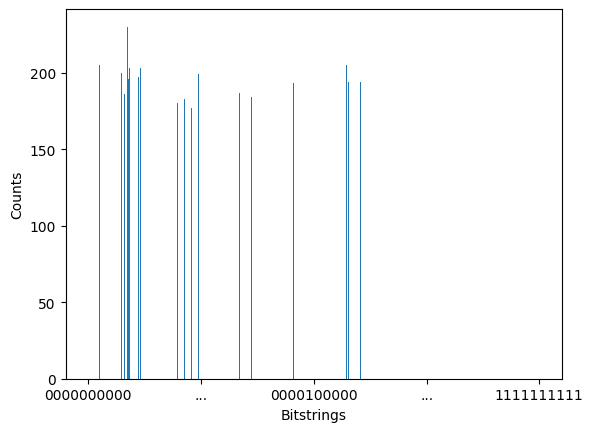}
    \caption{Results produced using the high-level functionality when solving for constraints consisting of 8 network state profiles with a corresponding number of occurrences.}
    \label{fig:high_level_8_constraint_simulator_bitstring_counts}
\end{figure}

First, if we consider 8 network state profiles as constraints and attempt to recover the regulatory logic governing the expression of protein Fgf8 in the noiseless quantum simulator environment, we obtain the frequency distribution of solutions shown in Figure \ref{fig:high_level_8_constraint_simulator_bitstring_counts} and Table \ref{tab:high_level_8_constraint_simulator_bitstring_counts}. We then filter these solutions to select for those proposed state transition models that comply with all 8 data-imposed constraints and convert each of these bitstring outputs into the Boolean expression representation.  This conversion produces a total of 51 possible logical state transition models for Fgf8 satisfying the given constraints, all of which are valid for the given (sparse) dataset.  Though not the top-ranking solution, we do recover the exact original logic proposed by \cite{weidner_2024} among the most probable (2.04\% occurrence) (model 14). Nonetheless, many other possible statements are also capable of satisfying our sparse constraints, with these occurring with individual probabilities of ~2\%.

\begin{table*}[ht]
\centering
\caption{\textnormal{State transition logic expressions obtained for the regulation of Fgf8 using 8 network state profiles with the implementation leveraging high-level functions, ranked by probability of occurrence (\%), with interpreted expressions.}}
\begin{tabular}{|c|c|c|l|}
\hline
\textbf{Position} & \textbf{Bitstring} & \textbf{Probability (\%)} & \textbf{Interpreted Expression} \\
\hline
1 & 0101001001 & 2.26 & Fgf8 $\land$ $\lnot$Emx2 $\land$ Pax6 $\land$ COUP-TFI \\
\hline
\multicolumn{4}{|c|}{\dots} \\
\hline
\textit{14} & 1101001001 & 2.04 & Fgf8 $\land$ $\lnot$Emx2 $\land$ Sp8 \\
\hline
\multicolumn{4}{|c|}{\dots} \\
\hline
51 & 0100101010 & 1.61 & $\lnot$Emx2 $\land$ COUP-TFI \\
\hline
\end{tabular}
\label{tab:high_level_8_constraint_simulator_bitstring_counts}
\end{table*}

\begin{figure}[H]
    \centering
    \includegraphics[width=1\linewidth]{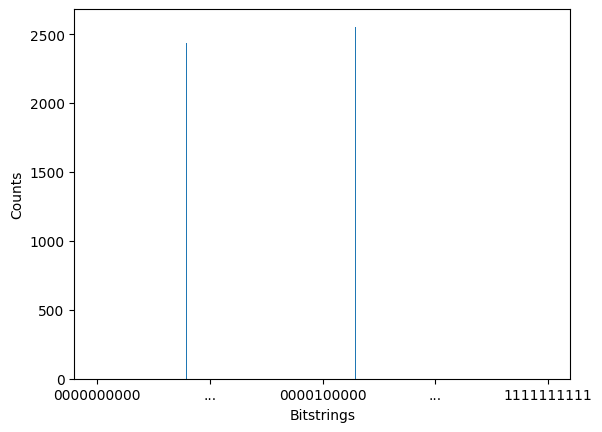}
    \caption{Resulting logic expression in bitstring form obtained by solving with high-level functions to satisfy 16 network state constraints, with corresponding number of occurrences.}
    \label{fig:high_level_16_constraint_simulator_bitstring_counts}
\end{figure}

\begin{table}[ht]
\centering
\caption{\textnormal{Bitstring solutions obtained using the circuit leveraging high-level functions for 16 network state constraints, ranked by probability, with interpreted expressions}}
\begin{tabular}{|c|c|c|l|}
\hline
\textbf{Position} & \textbf{Bitstring} & \textbf{Probability (\%)} & \textbf{Interpreted Expression} \\
\hline
1 & 1101001001 & 25.54 & Fgf8 $\land$ $\lnot$Emx2 $\land$ Sp8 \\
\hline
2 & 1101001101 & 25.27 & Fgf8 $\land$ $\lnot$Emx2 $\land$ Sp8 \\
\hline
3 & 1101001000 & 24.80 & Fgf8 $\land$ $\lnot$Emx2 $\land$ Sp8 \\
\hline
4 & 1101001100 & 24.39 & Fgf8 $\land$ $\lnot$Emx2 $\land$ Sp8 \\
\hline
\end{tabular}
\label{tab:high_level_16_constraint_simulator_bitstring_counts}
\end{table}

If we double the number of constraints to 16 simulated network state profiles, the number of competing solutions that equally satisfy the data decreases to only 4 valid bitstrings as seen in Figure \ref{fig:high_level_16_constraint_simulator_bitstring_counts} and Table \ref{tab:high_level_16_constraint_simulator_bitstring_counts}.  When we convert these from their bitstring representation into corresponding logic statements, we find that all 4 bitstrings represent exactly the same expression, specifically the original (i.e. true) Fgf8 state transition logic. Recall from the general parametrization in \ref{sec:assump} that each term is either included or excluded from the AND operation and can also be inverted (NOT) before inclusion. Accordingly, even when a term is excluded from the AND operation, it may be excluded in its original form or inverted (NOT). While this does not affect the correctness of the model, we will always get \(2k\) equivalent expressions where \(k\) the number of terms is excluded in the current parametrization.

\begin{table*}[t]
\centering
\caption{\textnormal{Solution logic statements with probability of occurrence obtained on IBM’s Eagle r3 quantum hardware using the low-level function manual implementation based on 8 network state profiles as constraints.}}
\begin{tabular}{|c|c|c|l|}
\hline
\textbf{Position} & \textbf{Bitstring} & \textbf{Probability (\%)} & \textbf{Interpreted Expression} \\
\hline
1 & 0001011101 & 0.33 & Sp8(t) \\
2 & 0110100101 & 0.31 & Emx2 $\land$ $\lnot$Pax6 $\land$ $\lnot$COUP-TFI \\
3 & 0001111101 & 0.30 & Sp8 $\land$ $\lnot$COUP-TFI \\
4 & 0011101101 & 0.29 & $\lnot$Pax6 $\land$ Sp8 $\land$ $\lnot$COUP-TFI \\
5 & 1101011101 & 0.29 & $\lnot$Fgf8 $\land$ $\lnot$Emx2 $\land$ Sp8 \\
6 & 1101010101 & 0.29 & $\lnot$Fgf8 $\land$ Emx2 $\land$ Sp8 \\
7 & 1101110101 & 0.28 & $\lnot$Fgf8 $\land$ Emx2 $\land$ Sp8 $\land$ $\lnot$COUP-TFI \\
8 & 1101001101 & 0.28 & \textcolor{softgreen}{Fgf8 $\land$ $\lnot$Emx2 $\land$ Sp8} \\
9 & 1011110101 & 0.27 & $\lnot$Fgf8 $\land$ Pax6 $\land$ Sp8 $\land$ $\lnot$COUP-TFI \\
10 & 1111111001 & 0.27 & $\lnot$Fgf8 $\land$ $\lnot$Emx2 $\land$ Pax6 $\land$ Sp8 $\land$ $\lnot$COUP-TFI \\
11 & 1000111101 & 0.27 & $\lnot$Fgf8 $\land$ $\lnot$COUP-TFI \\
12 & 1011110101 & 0.27 & $\lnot$Fgf8 $\land$ $\lnot$Pax6 $\land$ Sp8 $\land$ $\lnot$COUP-TFI \\
\hline
\multicolumn{4}{|c|}{\dots} \\
\hline
\end{tabular}
\label{tab:qpu_appproach1_8_constraints}
\end{table*}

\begin{table*}[t]
\centering
\caption{\textnormal{Solution logic statements with probability of occurrence obtained on IBM’s Eagle r3 quantum hardware using the high-level function implementation based on 8 network state profiles as constraints.}}
\begin{tabular}{|c|c|c|l|}
\hline
\textbf{Position} & \textbf{Bitstring} & \textbf{Probability (\%)} & \textbf{Interpreted Expression} \\
\hline
1 & 0001011100 & 0.41 & Sp8 $\land$ COUP-TFI \\
2 & 1010000100 & 0.37 & Fgf8 $\land$ $\lnot$Pax6 \\
3 & 0100101100 & 0.37 & \textcolor{softgreen}{$\lnot$Emx2 $\land$ COUP-TFI} \\
4 & 1001100100 & 0.37 & \textcolor{softgreen}{Fgf8 $\land$ Sp8 $\land$ COUP-TFI} \\
5 & 1101101100 & 0.36 & \textcolor{softgreen}{Fgf8 $\land$ $\lnot$Emx2 $\land$ Sp8 $\land$ COUP-TFI} \\
6 & 0000101100 & 0.36 & COUP-TFI \\
7 & 1110100100 & 0.36 & Fgf8 $\land$ Emx2 $\land$ $\lnot$Pax6 $\land$ COUP-TFI \\
8 & 0011100100 & 0.35 & $\lnot$Pax6 $\land$ Sp8 $\land$ COUP-TFI \\
9 & 0110000100 & 0.35 & Emx2 $\land$ $\lnot$Pax6 \\
10 & 1101010110 & 0.35 & \textcolor{softgreen}{Fgf8 $\land$ $\lnot$Emx2 $\land$ COUP-TFI} \\
11 & 1010100100 & 0.35 & Fgf8(t)$\land$$\lnot$Pax6(t)$/\land$COUP-TFI(t) \\
12 & 0100001100 & 0.34 & $\lnot$Emx2 \\
\hline
\multicolumn{4}{|c|}{\dots} \\
\hline
\end{tabular}
\label{tab:qpu_appproach2_8_constraints}
\end{table*}

\subsection{Execution on Real-World NISQ Hardware}

\begin{figure}[H]
    \centering
    \includegraphics[width=1\linewidth]{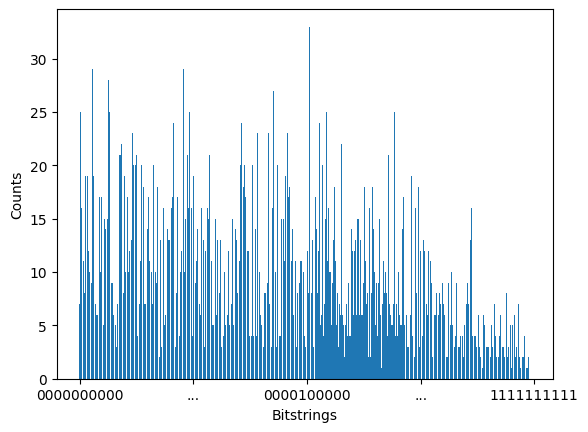}
    \caption{Bitstring solutions with number of occurrences obtained on IBM’s Eagle r3 quantum hardware using the low-level function manual implementation with 8 network state profiles as constraints. }
    \label{fig:qpu_appproach1_8_constraints}
\end{figure}

\begin{figure}[H]
    \centering
    \includegraphics[width=1\linewidth]{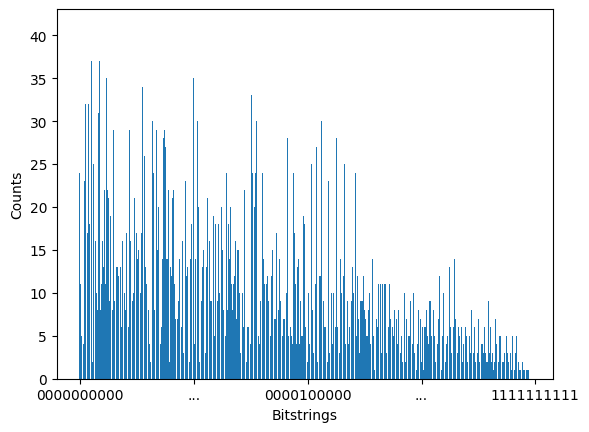}
    \caption{Bitstring solutions with number of occurrences obtained on IBM’s Eagle r3 quantum hardware using the high-level function implementation with 8 network state profiles as constraints.}
    \label{fig:qpu_appproach2_8_constraints}
\end{figure}

In the previous section, our multiple experiments verified that our algorithm can recover the true state transition logic for protein Fgf8 in an ideal noiseless simulated quantum environment. However, the key advantage of quantum computing resides in the potential for an exponential speedup offered by a sufficiently large system of noise-corrected physical qubits, at a scale exceeding classical simulation of such systems. We conducted numerical experiments using both low-level and high-level function implementations of our circuits deployed on IBM’s Eagle r3 quantum chip with 10,000 shots to characterize how real-world noise might affect our results. Starting with our benchmark problem using 8 network state profiles as constraints we obtain the outputs shown in Figure \ref{fig:qpu_appproach1_8_constraints} and Figure \ref{fig:qpu_appproach2_8_constraints} respectively. S

Note that after transpilation to the IBM quantum hardware language, our circuits designed with low-level and high-level functions now have a size 100568 and 226911, respectively, as reported by the Qiskit \textit{passmanager.run(qc).size()} routine. This final circuit size includes the oracle and diffuser executed iteratively \(m=3\) (8 constraint) and \(m=12\) (16 constraint) times. In each case, this was estimated as the optimal number iterations supporting proper convergence of Grover’s algorithm as described in section \ref{sec:optimal_iteration_number}. We immediately notice the relative probability of each top-ranking solution being much lower, and the output being much noisier, when compared to the simulated quantum environment (Table \ref{tab:qpu_appproach1_8_constraints}, Table \ref{tab:qpu_appproach2_8_constraints}). Green highlighting indicates a correct model as determined by classical methods; normal text indicates an incorrect model with respect to the given constraints.

Though both approaches produce noise-induced “incorrect” logic models with high rates of occurrence, the implementation using high-level functions appears to support better noise resilience. Specifically, the top-ranking solutions obtained with this implementation generally show a higher relative probability occurrence (0.41\% vs 0.31\% for top position) with a better contrast separating outputs. 

\section{Discussion}

In this work, we present promising results showing that it is possible to harness the scalability of quantum computing to reverse engineer the regulatory logic of a biological network.  This a very central problem as we begin to understanding the importance of network topology and dynamics in contributing to persistent illness and treatment resistance.  Using the example of a well-documented regulatory circuit describing the control of local developmental cues in the mammalian cortex, we recover the logic rules governing its behavior by formulating a Boolean satisfiability problem that can be deployed as a quantum circuit and solved using Grover’s algorithm. As mentioned previously, the recovery of feasible solutions from very sparse data is of broad interest in biomedical research. Our implementation exactly recovered the true logic statement for the specific protein of interest as well as a number of competing models that supported the few network states used as constraints.  An exhaustive list of feasible solutions was generated by conventional means and these exactly matched our quantum results obtained with the noise-free quantum simulator. While the formulation of this binary constraint satisfaction problem as two quantum circuit variants cannot and does not fundamentally resolve the indeterminacy of the sparse data problem (i.e. 8 data points), it is interesting to observe that the migration to the actual NISQ hardware and the use of low level functions, both contributors to increased noise, appear to support a broader search of the solutions space resulting in a better ranking of the true logic statement. The contribution of noise in improving search results is not unfamiliar.  Indeed, the noise inherent to an NISQ environment may confer some of the same advantages of a stochastic local search \cite{mengshoel_2008} but where this noise is embedded directly into the parameter space \cite{plappert_2017}.  Indeed, noise as a feature to be leveraged rather than simply mitigated has been gaining attention in the quantum community \cite{abdolazimi_2024,kashif_2024}.

In terms of execution performance, circuit design optimization using high-level functions, while promoting noise mitigation, resulted not only in less parsimonious solutions and poorer ranking but also performed less efficiently. While runtime is heavily dependent on the QPU technology, recovering logic statements satisfying 8 network state constraints on the IBM Eagle r3 system required a typical runtime of $\sim$3m:00s using the high-level function design, compared with $\sim$1:00s using low-level functions. Moreover, our implementation based on low-level functions resulted in a lower instruction count on QPU (100568 vs 226911), likely requiring higher qubit usage and associated increases in noise. While more susceptible to noise, our use of low-level functions would also be more amenable to broader platform-specific tuning. In contrast, the oracle generated using high-level functions would be much less flexible.

Despite the transportability of this simple binary logic problem to a much broader range of use cases, this analysis remains preliminary, and work is ongoing to investigate the robustness of this approach to the choice of initial data set as well as its scalability with increasing network complexity of the network and the corresponding increase in the number of constraints. In the current problem formulation, a larger number of samples translates into increasing gate depth.  Indeed, given the limitations of current NISQ hardware gate depths in the order of a few thousand approach a threshold beyond which the output would consist primarily of noise. Mitigation of decoherence \cite{jiang_2020} and optimization of resource usage as with increasing gate depth are active areas of research with optimization strategies in quantum machine learning ML in large data \cite{salehi_2022} being especially important as we continue to evaluate this aspect.  It is important to recall however that we are especially focused on the identification of network logic in data poor situations making our primary challenge that of network complexity. In this regard, the scale of the reverse engineering problem in this case is directly dependent on the in-degree or number of upstream regulators of a given target node (e.g. Fgf8) since the transition logic statement for each target may be resolved individually.  In this example problem the target Emx2 presented with the highest in-degree, namely 5 upstream regulators.  Though the average in-degree is typically $<8$ \cite{jiang_2020}, a small fraction of highly central target genes in transcriptional regulatory networks may be mediated by as many as 1,000 upstream transcription factors \cite{albert_2005}. While new quantum chip architectures emerge at a rapid pace, a possible approach for deployment on current hardware could consist in simply sub-dividing a highly regulated gene into smaller functional subunits.  In addition to the moderate degree of connectivity, the parameterization used to model this small 5-protein network was relatively simple, consisting of a product of candidate state variables with an optional inversion. Systems with more complex network dynamics might be addressed using additional parameters and would in turn offer a better test of this technology. For example, instead of the simple logic \(A \land B \land C \land D…\) one could allow for \((A \lor B) \land C\ \land D\) or similar. One could also allow for a broader variety of operations, such as NAND, XOR, etc... to describe more complex state transition logic. Such increases in complexity may better illustrate the benefit over classical algorithms and better delineate the threshold of “quantum advantage” in problems such as these. 

Finally, the scope of this paper did not include additional circuit optimization to a specific QPU’s topology, nor were the core circuit operations optimized significantly. Future work would involve such hardware-specific optimizations. Moreover, while we did not conduct extensive comparisons with performance classical computational methods in this analysis it is important to recall that the primary focus of this first investigation was feasibility. Indeed, the main objective of this analysis was to explore the practicality in the biomedical field of applying quantum computing to the reverse engineering of immune response networks at a scale that is immediately useful in advancing the network-informed design of immune interventions, both preventative and restorative.  In exploring this, we found that noise inherent to quantum platforms may promote a more pro-active exploration of the optimization space and deliver better answers. Even in a worst-case scenario where quantum performance were equivalent, the potential for more biologically plausible solutions is without question a highly impactful contribution.

\section{Acknowledgments}
This work was supported by the University of Saskatchewan’s Centre for Quantum Topology and Its Applications (quanTA) and by the Vaccine and Infectious Disease Organization (VIDO). VIDO receives operational funding from the Canada Foundation for Innovation (CFI) through the Major Science Initiatives Fund and from the Government of Saskatchewan through Innovation Saskatchewan and the Ministry of Agriculture. The quanTA Centre’s high-performance computational work has been advanced through a CFI John R. Evans Leaders Fund grant while access to quantum computing resources through IBM Quantum and PINQ2 has been made possible by a PrairiesCan Regional Innovation Ecosystems (RIE) contract (both awarded to SR).  We give special thanks to members of IBM Quantum, in particular Nick Bronn, Alexandre Choquette, Julien Chosson, Genya Crossman, Paul Gulyas, and Sean Wagner, as well as to Marie-Eve Boulanger, Eric Capelle, Gael Humbert, Maxime Marchand, Marie-Claude Messier, Christophe Pere, and Mathieu Thery of PINQ2, for their tremendous help in facilitating our quantum hardware access and helping us to make the most of it scientifically. We thank the University of Saskatchewan Advanced Research Computing (ARC) team for their efforts and responsiveness in creating an excellent local environment for this computational work. This article is submitted with the permission of the Director of VIDO.

\end{document}